\documentclass[twocolumn,showkeys,aps,prb,showpacs]{revtex4-1}
\UseRawInputEncoding
\usepackage{graphicx}
\usepackage[CJKbookmarks,dvipdfm,colorlinks,linkcolor=red,citecolor=blue]{hyperref}

\begin{document}

\title{Proposal for  valleytronic materials: ferrovalley metal and valley gapless semiconductor}

\author{San-Dong Guo$^{1}$,  Yu-Ling Tao$^{1}$,  Guang-Zhao Wang$^{2}$, Shaobo Chen$^3$ and Yee Sin Ang$^{4}$}
\affiliation{$^1$School of Electronic Engineering, Xi'an University of Posts and Telecommunications, Xi'an 710121, China}
\affiliation{$^2$Key Laboratory of Extraordinary Bond Engineering and Advanced Materials Technology of Chongqing, School of Electronic Information Engineering, Yangtze Normal University, Chongqing 408100, China}
\affiliation{$^3$College of Electronic and Information Engineering, Anshun University, Anshun 561000, People's Republic of China}
\affiliation{$^4$Science, Mathematics and Technology (SMT), Singapore University of Technology and Design (SUTD), 8 Somapah Road, Singapore 487372, Singapore}
\begin{abstract}
Valleytronic materials can provide new degrees
of freedom to future electronic devices. In this work, the concepts of the ferrovalley metal (FVM) and valley gapless semiconductor (VGS) are proposed, which can be achieved in valleytronic
bilayer systems by electric-field tuning, where the interaction between out-of-plane ferroelectricity and A-type antiferromagnetism can induce  layer-polarized anomalous valley Hall (LP-AVH) effect.
 The K and -K valleys of FVM are both metallic, and  electron and hole carriers simultaneously exist. In the extreme case, the FVM can become VGS by analogizing spin gapless semiconductor (SGS).
 Moreover, it is proposed that the valley splitting enhancement and valley polarization reversal can be achieved by electric field in valleytronic
bilayer systems.
  Taking the bilayer $\mathrm{RuBr_2}$ as an example, our proposal  is
confirmed by the first-principle calculations. The FVM and VGS can be achieved in  bilayer $\mathrm{RuBr_2}$ by applying electric field. With  appropriate  electric field range, increasing  electric field can enhance valley splitting, and the valley polarization can be reversed by flipping electric field direction. To effectively tune valley properties by electric field in bilayer systems, the parent monolayer should possess out-of-plane magnetization, and have large valley splitting.
 Our  results shed light on the
possible role of electric field  in tuning valleytronic  bilayer systems, and  provide a  way to design the ferrovalley-related material  by electric field.

\end{abstract}
\keywords{Valleytronics, Electric field, Bilayer ~~~~~~~~~~~~~~~~~~~~~~~Email:sandongyuwang@163.com}

\maketitle

\section{Introduction}
\begin{figure}
  \includegraphics[width=8cm]{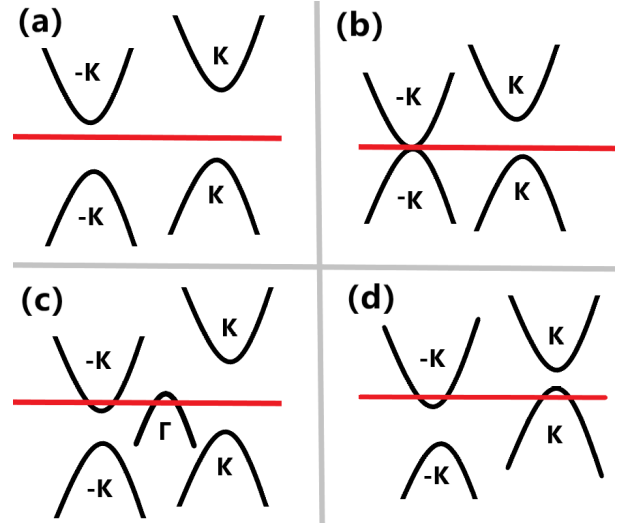}
\caption{(Color online) Schematic diagrams of energy bands of ferrovalley semiconductor (a), half-valley-metal (b),  quasi-half-valley-metal (c) and  ferrovalley metal (d). The horizontal red lines mean Fermi level.}\label{v-s}
\end{figure}

Manipulating different degrees of freedom of electrons plays a key role in building modern electronic devices.
The valley pseudospin is one of the emerging degrees of freedom beyond charge and
spin of carriers\cite{lv1,lv2}.  In crystalline
solids, valley is characterized by a local energy extreme for both the conduction  or valence band. For many $\mathrm{MoS_2}$-like monolayers,
their conduction band  minimum (CBM)  and valence
band maximum (VBM) are located in two inequivalent momenta -K and K, which  constitute a binary valley index\cite{lv1,q8-1,q8-2,q8-3,q9-1,q9-2,q9-3,q9-4}. The intensive efforts have been made
to manipulate the valley pseudospin, and a well-known field is established and  called valleytronics\cite{q9-5}.
The main challenge for valleytronics lies in inducing valley polarization. For systems with time-reversal symmetry, the optical pumping, magnetic field, magnetic
substrates and  magnetic doping have been proposed to  generate valley-polarized states\cite{q8-1,q8-2,q8-3,q9-1,q9-2,q9-3,q9-4,ar2,ar3}. However,  these methods have some disadvantages.  The magnetic
substrates and  magnetic doping  destroy  intrinsic  energy band structures and crystal structures. The optical pumping and  magnetic field limit the generation of
purely valley-polarized states. The concepts of ferrovalley semiconductor (FVS) [\autoref{v-s} (a)] and half-valley metal (HVM) [\autoref{v-s} (b)] with intrinsic spontaneous valley polarization have been proposed\cite{q10,q10-1}, which have been predicted in many two-dimensional (2D)   ferromagnets\cite{lv3,lv4,lv5,lv6,lv7,lv8,lv9,lv10,lv11,lv12,lv13,lv14,lv14-1,lv14-2}. Recently, we have proposed   possible electronic state quasi-half-valley-metal (QHVM) [\autoref{v-s} (c)], which contains electron and hole carriers with only a type of carriers being valley polarized\cite{lv15}.

\begin{figure}
  \includegraphics[width=8cm]{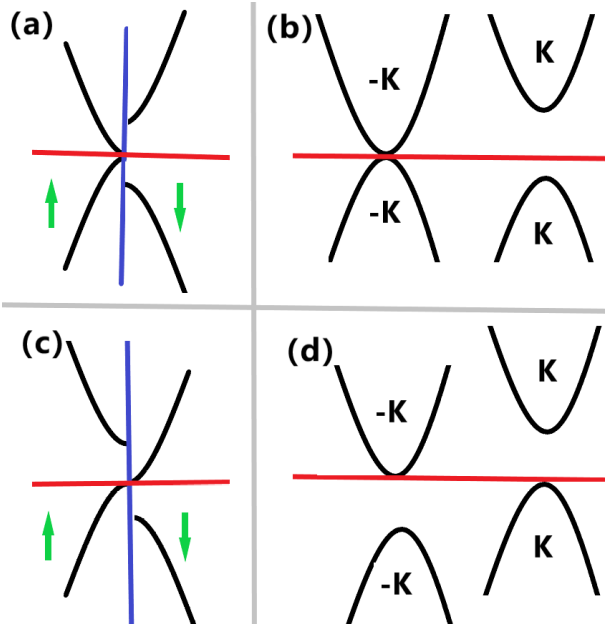}
\caption{(Color online)  Schematic diagram of the analogy between spin gapless semiconductor (a, c) and valley gapless semiconductor (b, d).
   The spin-up/spin-dn is equivalent to -K/K valley. The green arrows mean spin, and the horizontal red lines mean Fermi level.}\label{v-s-c}
\end{figure}

In analogy to ferromagnetic metal in spintronics, we propose the concept of ferrovalley metal (FVM) [\autoref{v-s} (d)], where the K and -K valleys are both metallic. For FVM,  electron and hole carriers simultaneously exist, and the Fermi level slightly touches -K and K valleys (CBM and VBM).
The concept of the spin gapless semiconductor (SGS) [\autoref{v-s-c} (a) and (c)],  where both electron and hole can be fully spin
polarized, has been proposed in spintronics\cite{sgs}. By analogizing SGS,  the concept of the valley gapless semiconductor (VGS) is proposed in this work. There are two
possible band structure configurations with valley  gapless
features as illustrated in \autoref{v-s-c} (b) and (d).  For the first
case [\autoref{v-s-c} (b)], one valley is gapless, while the other
valley is semiconducting. In fact, the first case is HVM, which has been proposed in ref\cite{q10-1}.
In the second case [\autoref{v-s-c} (d)],
there is a gap between the conduction and valence bands
for both the -K and K valleys, while there is
no gap between -K valley in the valence band
and K valley in the conduction band (or K valley in the valence band
and -K valley in the conduction band), which is named as
VGS-2. The second case is the extreme case of FVM, where the Fermi level exactly touches -K and K valleys (CBM and VBM).
The schematic diagrams of the analogy between SGS and VGS  are plotted in \autoref{v-s-c}.

\begin{figure}
  \includegraphics[width=7cm]{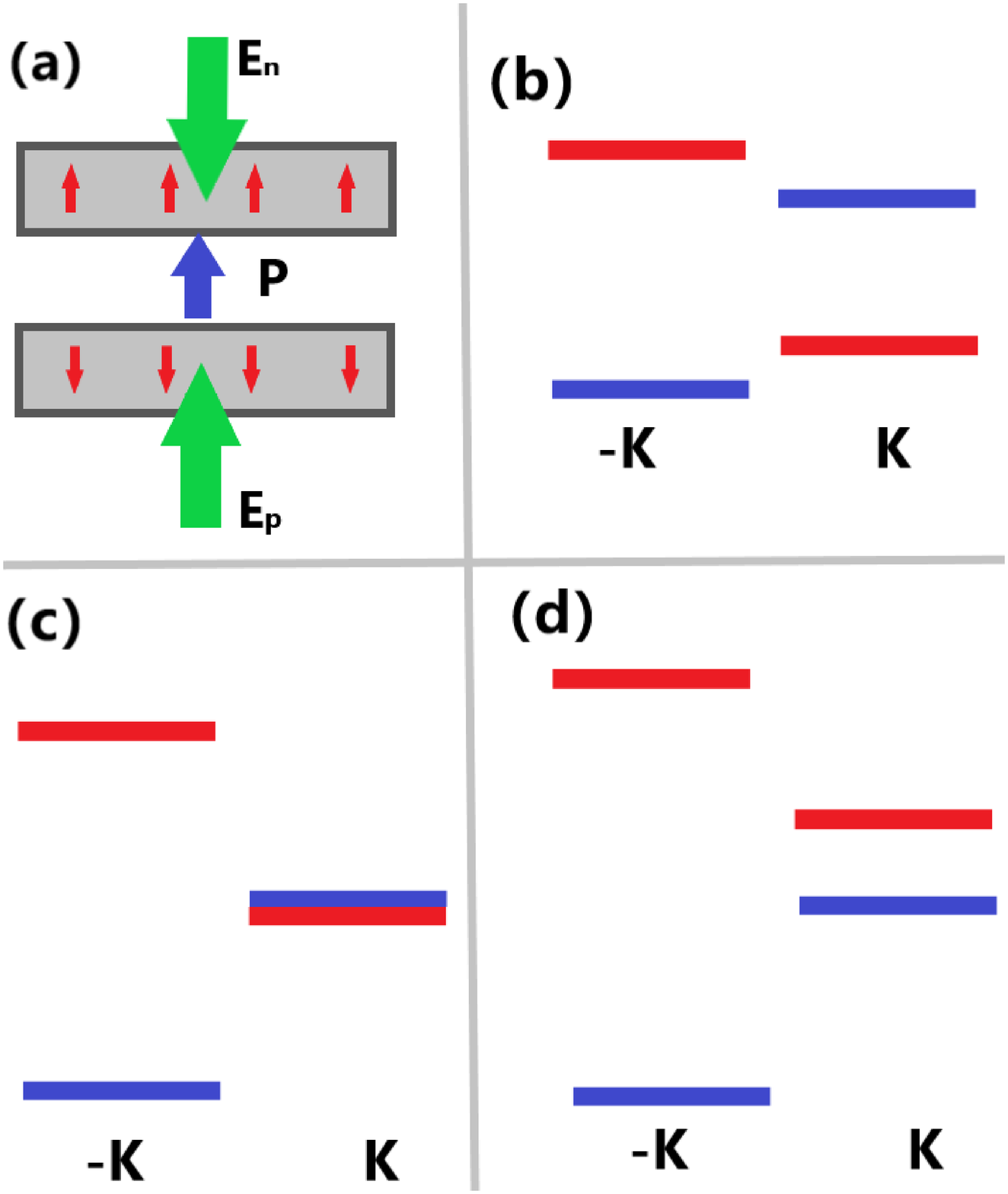}
  \caption{(Color online)(a): The schematic diagram of  a bilayer lattice with AB pattern, and the red, blue and green arrows represents spin,  electric polarization $P$ and external electric field $E_p$ (positive $z$ direction) and $E_n$ (negative $z$ direction).  The energy level of  -K and K valleys without (b) and with  external electric field $E_p$=$E_p^c$ (c) and $E_p$$>$$E_p^c$ (d). The red (blue) energy levels are from dn-layer (up-layer).}\label{st}
\end{figure}
\begin{figure*}
  \includegraphics[width=16cm]{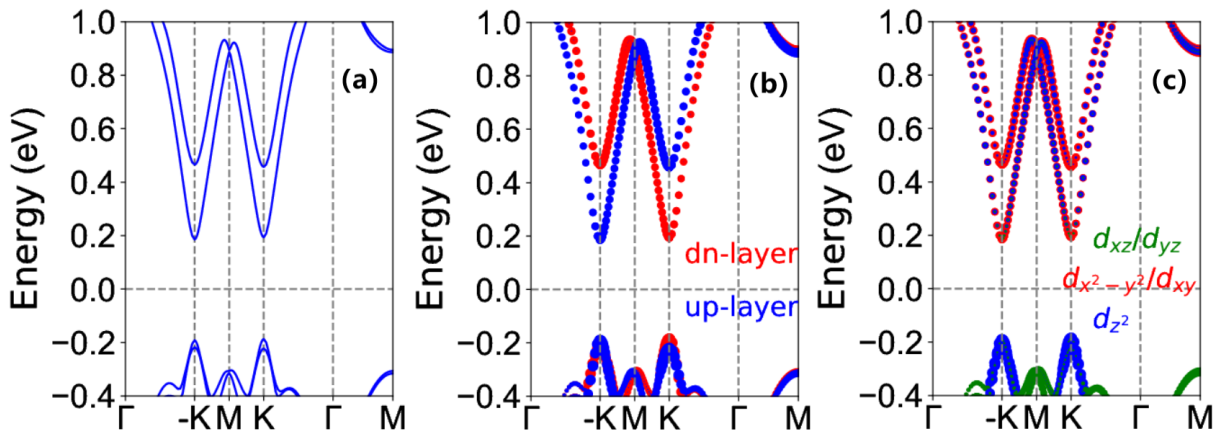}
  \caption{(Color online)(a): The band structures of AB-stacked bilayer $\mathrm{RuBr_2}$,  including layer- (b) and  Ru-$d$-orbital (c) characters energy band structures. }\label{band}
\end{figure*}

It is difficult to find these materials of FVM and VGS-2 in simple compounds.
Recently, layer-polarized anomalous Hall  effect in
valleytronic van der Waals bilayers  by interlayer sliding has been proposed  from
2D systems with spontaneous valley polarization\cite{lv16}.  The interaction between the
out-of-plane ferroelectricity and A-type antiferromagnetism allows the
realization of layer-polarized anomalous valley Hall (LP-AVH) effect.  The ferroelectric switching can induce reversed sign change of valley polarization . The out-of-plane ferroelectricity polarization is equivalent to an electric field\cite{ar1}, so an external electric field can be used to tune layer valley polarization, which provides possibility to achieve FVM and VGS-2.

As shown in \autoref{st} (a), a AB-stacked bilayer lattice from a 2D system with main spontaneous valley polarization in the conduction bands ($d_{x^2-y^2}$+$d_{xy}$-dominated  orbitals) has positive  electric polarization, and layer spontaneous valley polarization can be observed (\autoref{st} (b)).
Without out-of-plane electric polarization,  the energies of -K and K valleys from different layer   are coincident.
 By applying an out-of-plane electric polarization or   external electric field
penetrating from dn-layer  to up-layer  (defined as the positive field) (see \autoref{st} (a)),
the energy band from dn-layer is shifted toward high energy  with respect to one of up-layer, which leads to spontaneous valley polarization.
As long as  the valley splitting of pristine monolayer between -K and K valleys  is  big enough (This is larger than one caused by out-of-plane electric polarization in bilayer system.), the -K and K valleys of  bilayer system are from different layer  (see \autoref{st} (b)). When increasing   positive external electric field $E_p$, the energy levels at K valley from different layers will coincide at a critical electric field $E_p^c$  (see \autoref{st} (c)). When continuing  to increase $E_p$, the -K and K valleys of  bilayer system are from the same up-layer  (see \autoref{st} (d)).
 When an appropriate positive external electric field ($E_p$$<$$E_p^c$) is applied, the spontaneous valley polarization (valley splitting) should be  enhanced. However, when the external electric field is reversed (An appropriate negative external electric field ($E_n$) is applied.), the sign of valley polarization can  also be  reversed (The -K and K valleys along the levels from up-layer and dn-layer exchange each other.).
The above analysis also apply  to the valence bands ($d_{z^2}$-dominated  orbitals) with small spontaneous valley polarization, and the  critical electric field $E_p^v$ is very small. By applying an out-of-plane  external electric field
penetrating from dn-layer  to up-layer,
the energy band from dn-layer is shifted toward high energy  with respect to one of up-layer, which can realize VGS-2 and FVM.

 Here,  a concrete example of  bilayer $\mathrm{RuBr_2}$ is used to illustrate our idea.
Calculated results show that increasing electric field indeed can enhance valley splitting in bilayer $\mathrm{RuBr_2}$, and make K and -K valleys be from the same layer. The possible electronic states VGS-2 and  FVM can be achieved in  bilayer $\mathrm{RuBr_2}$ caused by electric filed.
 Our findings can be extended to other valleytronic  bilayers, and  tune their valley properties  by electric field.

The rest of the paper is organized as follows. In the next
section, we shall give our computational details and methods.
 In  the next  section,  we shall present crystal structure,  electronic structures  and  electric field effects on  physical properties  of  bilayer $\mathrm{RuBr_2}$. Finally, we shall give our discussion and conclusion.

\begin{figure*}
  \includegraphics[width=16cm]{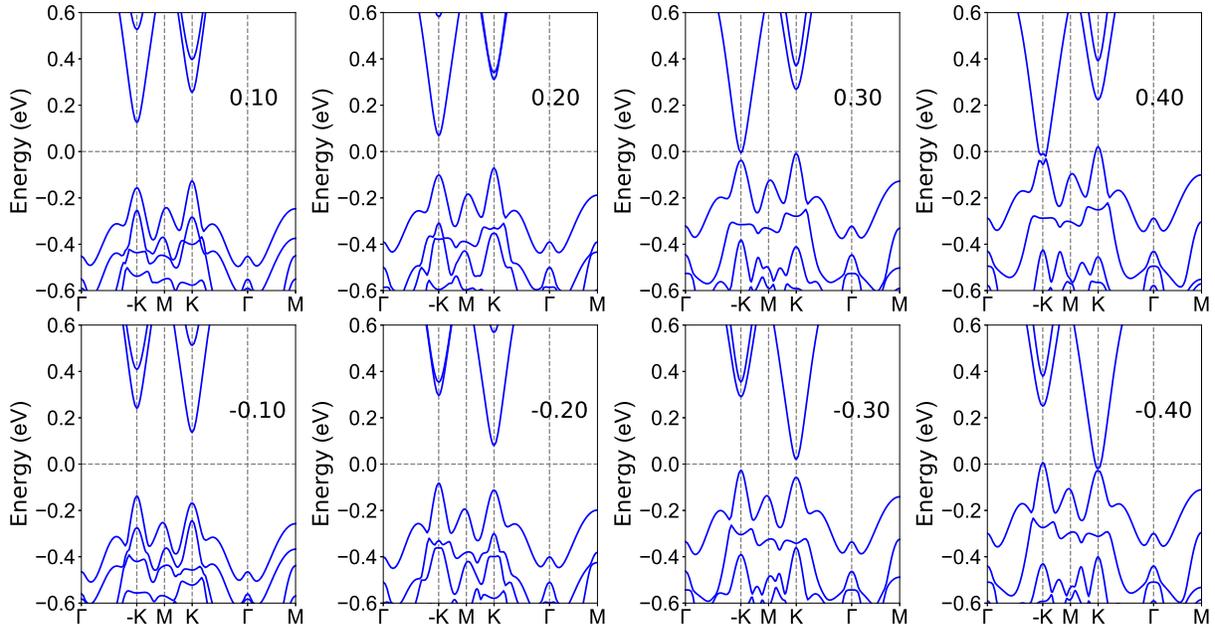}
  \caption{(Color online)The energy band structures of AB-stacked bilayer $\mathrm{RuBr_2}$ at representative $E$=$\pm$0.10, $\pm$0.20, $\pm$0.30 and $\pm$0.40 $\mathrm{V/{\AA}}$. }\label{band-1}
\end{figure*}

\begin{figure}
  \includegraphics[width=8cm]{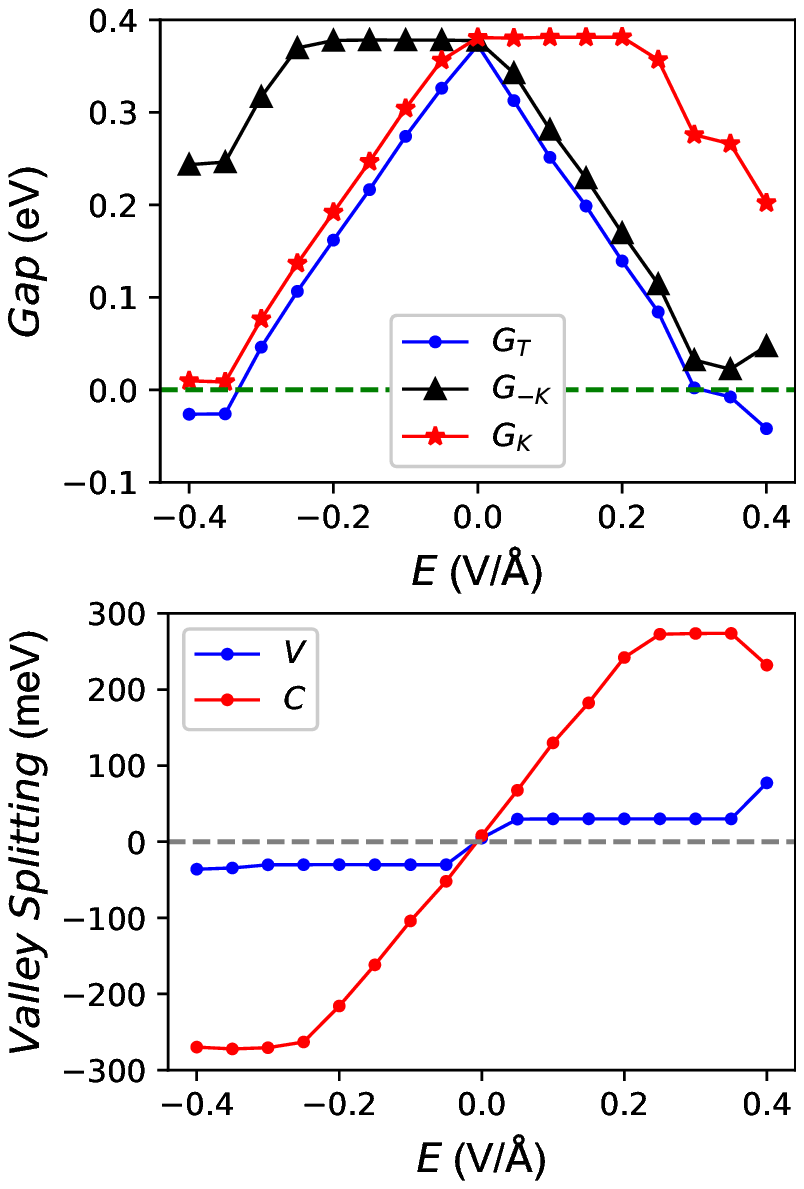}
  \caption{(Color online)The related  band gaps  (top panel)  and   valley splitting  for both valence and condition bands (bottom panel) of AB-stacked bilayer $\mathrm{RuBr_2}$ as a function of   $E$.  }\label{gap}
\end{figure}

\section{Computational detail}
Within density-functional
theory  (DFT)\cite{1},  the spin-polarized  calculations are carried out  by employing the projected
augmented wave method,  as implemented in VASP code\cite{pv1,pv2,pv3}.
We use the generalized gradient approximation of Perdew-Burke-Ernzerhof (PBE-GGA)\cite{pbe}  as exchange-correlation functional.
To consider on-site Coulomb correlation of Ru atoms,  the GGA+$U$   method   in terms of  the on-site Coulomb
interaction of $U$$=$ 2.5 eV\cite{lv14,lv14-1} is used within  the rotationally invariant approach proposed by Dudarev et al\cite{u}.
To attain accurate results, we use the energy cut-off of 500 eV,  total energy  convergence criterion of  $10^{-7}$ eV and  force
convergence criteria of less than 0.001 $\mathrm{eV.{\AA}^{-1}}$ on each atom.
To avoid the interactions
between the neighboring slabs, a vacuum space of more than 20 $\mathrm{{\AA}}$ is used. The dispersion-corrected DFT-D3 method\cite{dft3} is adopted to describe the van der Waals
interactions between individual layers.
 The $\Gamma$-centered 18 $\times$18$\times$1 k-point meshs  in  the Brillouin zone (BZ) are used for structure optimization and electronic structures calculations.
 The spin-orbital coupling (SOC) effect is explicitly included to investigate magnetic anisotropy energy (MAE) and  electronic  properties of   bilayer $\mathrm{RuBr_2}$.
The Berry curvatures are calculated  directly from  wave functions  based on Fukui's
method\cite{bm} by using  VASPBERRY code\cite{bm1,bm2}. Under an electric field, the atomic positions are relaxed. To easily meet  energy  convergence criterion,   the parameter DIPOL=0.5 0.5 0.5 is set, and  the  convergent charge density under small electric field gradually feeds to the calculations with large electric field.

\section{electronic structures}
The LP-AVH effect has been demonstrated in a series of  valleytronic
materials, such as  $\mathrm{VSi_2P_4}$, $\mathrm{VSi_2N_4}$, $\mathrm{FeCl_2}$, $\mathrm{RuBr_2}$ and VClBr\cite{lv16}.
To clearly demonstrate our previous analysis of electric field effects on valley polarization in valleytronic  bilayers (\autoref{st}), the parent monolayer should have large valley splitting. The previous works show that $\mathrm{RuBr_2}$ has very large valley splitting in the conduction bands or valence bands, which depends on the
electronic correlation strength or strain\cite{lv14,lv14-1}. Therefore, the $\mathrm{RuBr_2}$ monolayer is used to validate our proposal.

The $\mathrm{RuBr_2}$ monolayer consists Br-Ru-Br sandwich layer, and shares the same crystal structure with $\mathrm{MoS_2}$.  It has a hexagonal lattice
with the space group $P\bar{6}m2$, and  its inversion symmetry  is broken,  which along with ferromagnetic
(FM) ordering can give rise to ferrovalley features.
The $\mathrm{RuBr_2}$  shows a spontaneous valley splitting of 265
(31) meV in the conduction (valence) band edge at $U$$=$2.5 eV\cite{lv14}. Here, we only  construct AB-stacked bilayer $\mathrm{RuBr_2}$, which is plotted in FIG.1 of  electronic supplementary information (ESI).
The BA-stacked bilayer has the same results with AB-stacked case, when  the electric field and  sign of valley polarization
are simultaneously reversed.

 The  bilayer $\mathrm{RuBr_2}$  has the space group of $P3m1$, whose inversion
symmetry and horizontal mirror symmetry are broken. The optimized lattice constant of
bilayer $\mathrm{RuBr_2}$  is 3.72 $\mathrm{{\AA}}$, and the interlayer distance is 3.16  $\mathrm{{\AA}}$. The AB
and BA cases are energetically degenerate  with opposite
electric polarizations,  and  connect  each other  by interlayer sliding\cite{lv16}. The spontaneous out-of-plane electric polarization is  along positive $z$ direction in the AB-stacked bilayer.
To determine the ground state of bilayer $\mathrm{RuBr_2}$, the intralayer FM and interlayer FM, and intralayer FM and interlayer antiferromagnetic (AFM)  magnetic configurations are considered. Calculated results show that bilayer $\mathrm{RuBr_2}$ prefers A-type antiferromagnetism (intralayer FM and interlayer AFM orderings).
This A-type antiferromagnetism is 7.4 meV per Ru atom lower than that with
the FM interlayer exchange interaction.

The energy band structures of AB case are plotted in \autoref{band} along with layer-  and  Ru-$d$-orbital  characters ones.
The AB bilayer shows an indirect band gap of 0.373 eV, and  the VBM and CBM locate at the K  and -K points, respectively.
 It is clearly seen that the VBM (CBM) is from the dn-(up-)layer (\autoref{band} (b)), and the valleys are layer-locked  with spontaneous valley polarization.
The  valley splitting in the valence (conduction) bands are defined as: $\Delta E_V=E_V^K-E_V^{-K}$ ($\Delta E_C=E_C^K-E_C^{-K}$),   and the  calculated
value is 4.8 meV (8.10 meV). According to \autoref{band} (c), it is found that the valley splitting of conduction band of monolayer  $\mathrm{RuBr_2}$ is observable, while the valley splitting of valence band is very small, which is due to different distribution of Ru-$d$ orbitals.
The valley splitting $|\Delta E|$  can be expressed as\cite{v2,v3}: $|\Delta E|=|E^{K}-E^{-K}|=4\alpha$ ($\alpha$ is  the SOC-related constant), when $d_{x^2-y^2}$+$d_{xy}$  orbitals dominate  the K and -K valleys. If the -K and K valleys are mainly from $d_{z^2}$ orbitals, the valley splitting $|\Delta E|$ will become:$|\Delta E|=|E^{K}-E^{-K}|\approx0$.

Essentially, ferroelectricity polarization and electric field are equivalent to produce valley polarization in valleytronic  bilayer.
Here, the  electric field effects  on valley polarization in   bilayer $\mathrm{RuBr_2}$ are investigated.
 Firstly, we determine the magnetic ground state under the positive and negative electric field, and the energy differences (per Ru atom) between interlayer  FM and AFM ordering as a function of electric field $E$ are shown in FIG.2 of ESI.
Calculated results show that the interlayer  AFM state is always  ground state within  considered $E$ range, and applied  electric field can enhance the interlayer  AFM interaction.

 The energy band structures of  bilayer $\mathrm{RuBr_2}$ under representative  electric field $E$ are plotted in \autoref{band-1}, and the evolutions of related energy band gap  and the valley splitting for both valence and condition bands  as a function of $E$ are plotted in \autoref{gap}.
For increasing positive $E$, the global gap decreases, and  a semiconductor to metal transition is induced at $E$$=$0.30  $\mathrm{V/{\AA}}$.
By applying positive $E$, the energy band from dn-layer is shifted toward high energy  with respect to one of up-layer. When increasing   positive  $E$, the energy levels at K valley  from different layers in the conduction bands near Fermi level will coincide at about  $E$$=$0.20 $\mathrm{V/{\AA}}$.
For  $E$$>$0.20 $\mathrm{V/{\AA}}$,  the -K and K valleys of  bilayer system in the conduction bands are from the same up-layer, which can be clearly seen from FIG.3 of ESI (for example $E$$=$0.40 $\mathrm{V/{\AA}}$). For the valence bands, these phenomenons can also be observed, and the critical $E$ is very small, which is due to small valley splitting in the valence bands for monolayer  $\mathrm{RuBr_2}$ at $U$$=$2.5 eV.
With increasing   positive  $E$,  the gap of K valley firstly  remains almost unchanged, and then decreases. However, for the gap of -K valley, it decreases, and then increases at about  $E$$=$0.35 $\mathrm{V/{\AA}}$.
For  0  $\mathrm{V/{\AA}}$$<$$E$$<$0.25  $\mathrm{V/{\AA}}$, the valley splitting in the conduction bands increases with increasing positive $E$.
At  $E$$=$0.25  $\mathrm{V/{\AA}}$, the valley splitting of conduction bands reaches up to 273 meV, which is close to one (265 meV) of monolayer  $\mathrm{RuBr_2}$\cite{lv14}. The analysis above can also be applied to negative $E$ case.
The difference mainly includes two aspects: (1) the K and -K valleys exchange each other; (2) the negative $E$ firstly need to cancel out the small polarized electric field. When an appropriate positive electric field is reversed, the sign of valley polarization can  also be  reversed.

\begin{figure*}
  \includegraphics[width=12cm]{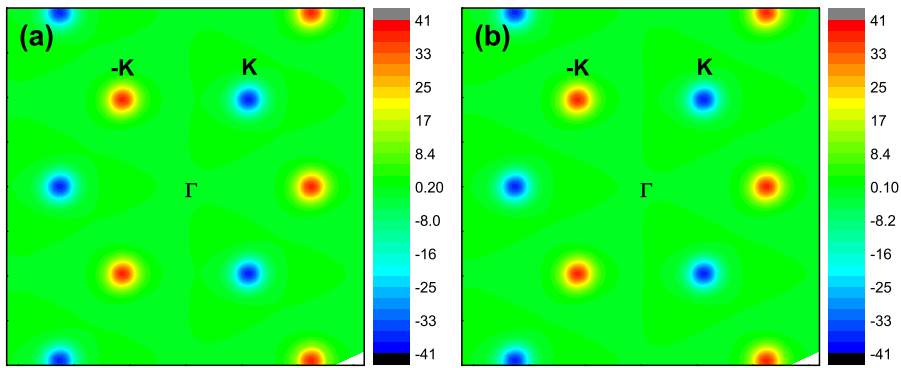}
\caption{(Color online)The Berry curvatures of  AB-stacked bilayer $\mathrm{RuBr_2}$ in BZ at representative $E$=0.10 $\mathrm{V/{\AA}}$ (a) and -0.10 $\mathrm{V/{\AA}}$ (b).}\label{be}
\end{figure*}

\autoref{be} present the calculated  Berry curvatures $\Omega(k)$ of
bilayer $\mathrm{RuBr_2}$ under $E$$=$$\pm$0.10   $\mathrm{V/{\AA}}$.
The valley splitting for the conduction (valence) band  is 130 (30.0) meV   and  104 (30.2) meV for  $+$0.10   $\mathrm{V/{\AA}}$ and  $-$0.10   $\mathrm{V/{\AA}}$, which are larger than one (8.1 meV (4.8 meV)) without applying $E$.
For  $+$0.10   $\mathrm{V/{\AA}}$ case,  the energy of K valley
is higher than one of -K valley.  The valley polarization can  be
switched by reversing the electric field direction from $+z$ to $-z$ direction.
For the two situations, we observe  opposite
signs of Berry curvature around -K and K valleys with the unequal magnitudes.   By reversing
the electric field, the magnitudes
of Berry curvature at -K and K valleys exchange  each
other,   but their signs remain unchanged.

Under an in-plane longitudinal  $E$, Bloch electrons at K and -K valleys will obtain anomalous velocity:$\upsilon\sim E\times\Omega(k)$\cite{qqq}.
An appropriate  doping
makes the Fermi level  fall between the -K and K
valleys.  With applied in-plane and out-of-plane  electric fields, the  Berry curvature forces
the  carriers to accumulate on one side of one layer of bilayer. When the out-of-plane  electric field is reversed, the  carriers  accumulate on one side of the other layer of bilayer. These give rise to LP-AVH effect.

 The concepts of FVS and HVM have been proposed, which can achieve spontaneous valley polarization\cite{q10,q10-1}.
 For FVS, the K and -K valleys are both insulating (\autoref{v-s} (a)). For HVM,  one of K and -K valleys is insulating, while the other is metallic (\autoref{v-s} (b)). The HVM can be realized by changing electron correlation strength or applying strain, which is just at one point, not a region of electron correlation strength or strain. Recently, the  QHVM is proposed (\autoref{v-s} (c)), where electron and hole carriers simultaneously exist with only a type of carrier being valley polarized (The Fermi level slightly touches CBM and VBM, and the carriers around $\Gamma$ possess
almost zero Berry curvatures.)\cite{lv15}.  Here, the FVM is proposed (\autoref{v-s} (d)), where the K and -K valleys are both metallic. For FVM,  electron and hole carriers simultaneously exist, and the Fermi level slightly touches CBM and VBM.
Under an in-plane longitudinal  $E$, the  Berry curvature forces
the electron carriers to accumulate on one side of sample, and hole ones to move toward the other side.
In the bilayer $\mathrm{RuBr_2}$, the FVM can be achieved by applying electric filed.
At  $E$$=$0.40 $\mathrm{V/{\AA}}$, the Fermi level slightly touches the -K valley of conduction bands and K valley of valence bands, which are from different layer (FIG.3 of ESI). The gaps of -K and K valleys are 47 meV and 202 meV, respectively.
Under an in-plane longitudinal  $E$, the electron and hole carriers accumulate on one side of different layer.
In the extreme case, the Fermi level touches the valley bottom
of  -K valley of conduction bands and the valley top of  K valley of valence bands  (\autoref{band-1} at $E$$=$0.30  $\mathrm{V/{\AA}}$), which can achieve VGS-2 (\autoref{v-s-c} (d)).

\section{Discussion and Conclusion}
For monolayer FVS with hexagonal symmetry, the spontaneous valley polarization depends on the magnetization direction\cite{lv3,lv4,lv5,lv6,lv7,lv8,lv9,lv10,lv11,lv12,lv13,lv14,lv14-1}. For out-of-plane magnetization, the monolayer FVS possess spontaneous valley polarization. However, for in-plane magnetization, no spontaneous valley polarization can be produced.
For bilayer system from parent monolayer with in-plane magnetization, the K and -K valleys are from the same layer.  By applying an out-of-plane electric polarization or   external electric field, the bilayer system has not spontaneous valley polarization. To confirm this, layer-characters energy band structures at $E$$=$$\pm$0.15   $\mathrm{V/{\AA}}$ are plotted in FIG.4 of ESI with in-plane magnetization. It is clearly seen that the K and -K valleys of both valence and conduction bands are from the same layer, and no spontaneous valley polarization can be observed.

To determine magnetization direction, we calculate  the MAE of bilayer $\mathrm{RuBr_2}$, which is  defined as the energy difference with the magnetization axis along in-plane and out-of-plane directions. The MAE as a function of $E$ is plotted in FIG.5 of ESI, which  indicates that bilayer $\mathrm{RuBr_2}$ favor in-plane magnetization orientation within considered $E$ range due to negative MAE. Therefore, bilayer $\mathrm{RuBr_2}$ intrinsically has no spontaneous valley polarization at $U$$=$2.5 eV. The previous work shows that the magnetic anisotropy direction of monolayer $\mathrm{RuBr_2}$ changes  from out-of-plane to in-plane one with  the critical $U$ value of 2.07 eV\cite{lv14}.  If  the real $U$ falls in  the range ($U$$<$2.07 eV), bilayer $\mathrm{RuBr_2}$ will possess spontaneous valley polarization.

Even though the real $U$ falls outside the range, the spontaneous valley polarization can be achieved by strain.
 By applying strain, the bandwidth can be modified, which
effectively controls the relative importance of electronic correlation. To reduce relative importance of electronic correlation, the compressive strain should be used, which equivalently reduces $U$ value. To demonstrate this point, $a/a_0$=0.95 biaxial strain is applied  on the bilayer $\mathrm{RuBr_2}$ with $U$$=$2.5 eV.
Calculated results show that strained bilayer $\mathrm{RuBr_2}$ prefers A-type antiferromagnetism, which  is 9.8 meV per Ru atom lower than that with
the FM interlayer exchange interaction. The calculated MAE is 909 $\mathrm{\mu eV}$/Ru, which  indicates that strained bilayer $\mathrm{RuBr_2}$ favor out-of-plane magnetization orientation. The energy band structures of bilayer $\mathrm{RuBr_2}$ is plotted in FIG.6 of ESI, and the  valley splitting in the valence (conduction) bands  is -14.6 meV (-3.0 meV).

In summary, we have demonstrated that the electric field  can effectively tune valley properties of bilayer $\mathrm{RuBr_2}$.
 The FVM and VGS-2 can be realized in  bilayer $\mathrm{RuBr_2}$ by electric field tuning. In addition, the  electric field can enhance the valley splitting of  bilayer $\mathrm{RuBr_2}$, and make the -K and K valleys be from the same layer.
We take bilayer $\mathrm{RuBr_2}$ as a concrete example, but  the analysis here can be readily extended
to other valleytronic van der Waals bilayers.
Our findings can   expand  understanding of  valleytronic van der Waals bilayers, and realize new valleytronic materials: FVM and VGS-2.
~~~~\\
~~~~\\
\textbf{SUPPLEMENTARY MATERIAL}
\\
See the supplementary material for  crystal structures; energy difference between FM and AFM and MAE as a function of $E$; the related energy band structures.

~~~~\\
~~~~\\
\textbf{Conflicts of interest}
\\
There are no conflicts to declare.

\begin{acknowledgments}
This work is supported by Natural Science Basis Research Plan in Shaanxi Province of China  (2021JM-456). We are grateful to Shanxi Supercomputing Center of China, and the calculations were performed on TianHe-2.
\end{acknowledgments}

\end{document}